\documentclass[preprint,aps,showpacs]{revtex4}
\usepackage[dvips]{graphicx}
\usepackage{color}
\usepackage{amssymb}
\usepackage{amsmath}
\usepackage{units}
\usepackage[latin1]{inputenc}

\begin{document}

\title{Determination of exchange constants of Heusler compounds by Brillouin light scattering spectroscopy: application to Co$_2$MnSi}

\author{J.~Hamrle%
\footnote{Corresponding author: J. Hamrle, email: hamrle@physik.uni-kl.de}%
, O.~Gaier, Seong-Gi Min, B.~Hillebrands}
\affiliation{Fachbereich Physik and Forschungszentrum OPTIMAS,
Technische Universit\"at Kaiserslautern,
Erwin-Schr\"odinger-Stra\ss e 56, D-67663 Kaiserslautern, Germany}

\author{Y.~Sakuraba}
\affiliation{Magnetic Materials Laboratory, Institute for
Materials Research (IMR), Tohoku University, Katahira 2-1-1,
Sendai 980-8577, Japan}

\author{Y.~Ando}
\affiliation{Department of Applied Physics, Graduate School of
Engineering, Tohoku University, Aoba-yama, Sendai 980-8579, Japan}

\begin{abstract}

Brillouin light scattering spectroscopy from so-called standing spin waves in thin magnetic films is often used to determine the magnetic exchange constant. 
The data analysis of the experimentally determined spin-wave modes requires an unambiguous assignment to the correct spin wave mode orders. Often additional investigations are needed to guarantee correct assignment.
This is particularly important in the case of Heusler compounds where values of the exchange constant vary substantially between different compounds. 
As a showcase, we report on the determination of the exchange constant (exchange stiffness constant) in Co$_2$MnSi, which is found to be $A=2.35\pm0.1$\,$\mu$erg/cm ($D=575\pm20$\,meV\,\AA$^2$), a value comparable to the value of the exchange constant of Co.
\end{abstract}


\maketitle

\section{Introduction}

The investigation of electron-electron interactions in
half-metallic ferromagnetic Heusler compounds is an important
issue in order to understand the strong temperature dependence of
spin polarization of these materials. One of the key parameters in
this context is the magnetic exchange constant (in the following simply referred to as exchange constant) which describes the strength of
the exchange interaction between two spins inside a ferromagnetic
system. Brillouin light scattering (BLS) spectroscopy from standing spin-waves in thin magnetic films is a
well-established technique for the study of exchange interaction
in various material systems \cite{hillebrands07,liu96,buch03}. 
However, the application of this
experimental technique for the determination of exchange constants
in thin films of Heusler compounds presents some difficulties, which, as will
be discussed in this article, are mostly related to an ambiguity in the assignment of the measured mode frequencies to the correct standing spin-wave mode orders. The goal of this article is to show
in detail how the values of exchange constants are determined from
the BLS spectra measured on thin films of Heusler compounds as well as to discuss the
difficulties in the extraction of the exchange constants from the
experimental data. For this purpose, we present BLS studies of Co$_2$MnSi
films in this article.

In the following, we briefly describe the investigated Co$_2$MnSi
films and the determination of the exchange constant by means of BLS. Thereafter, the
experimental results of BLS studies performed on Co$_2$MnSi thin
films are presented, and a procedure leading to the correct mode assignment is discussed.

\section{Experimental details}

The investigated Co$_2$MnSi films with thicknesses
$t=20$, 30, 40, 60 and \unit[80]{nm} were epitaxially grown on a
MgO(100) substrate covered by a \unit[40]{nm} thick Cr(100) buffer
layer. For the deposition of the Co$_2$MnSi layers, inductively
coupled plasma-assisted magnetron sputtering was employed. A
post-growth annealing at \unit[500]{$^\circ$C} provided Co$_2$MnSi films
with a predominant L2$_1$ order, which was confirmed by x-ray
diffraction (XRD) measurements. The films were
covered by a \unit[1.3]{nm} thick Al protective layer to prevent
sample oxidation.

The BLS measurements presented in this article were performed at room temperature in
the magneto-static surface mode geometry where the magnetic field
$\vec{H}$ is applied in the plane of the sample and perpendicular to the plane of
light incidence i.e.\ perpendicular to the transferred wave vector $q_\|$ of the detected
magnons. A diode pumped, frequency doubled Nd:YVO$_4$ laser with a
wavelength of $\lambda=532$~nm was used as a light source. A
detailed description of the BLS setup used in this work can be
found for example in Refs.~\cite{moc87, hil99}. The BLS spectra of
Co$_2$MnSi films with varying thickness $t$ were recorded both at
different values of the external magnetic field $\vec{H}$ and at
different angles of incidence of the probing light beam $\varphi$, i.e., the angle between
the direction of the incident laser beam and the film normal. The
former is necessary to confirm the magnonic origin of peaks,
whereas the latter allows for the detection of spin waves with
different transferred wave vectors $q_\|$
($q_\parallel=4\pi/\lambda\sin\varphi$). As will be shown
later on in this article, $q_\|$ dependent measurements of BLS
spectra are required for an unambiguous separation of the dipole
dominated magnetostatic surface wave, also called Damon-Eshbach
mode (DE), from the exchange dominated perpendicular standing spin
waves (PSSW).

The analytical expression of frequencies of DE and PSSW modes is presented e.g.\ in Refs.~\cite{dem01,gai08,kal86}. 
DE mode is
characterized by an exponential decay of the amplitude of the
dynamic magnetization \footnote{The dynamic magnetization is the difference between the static
magnetization $\vec{M}_0$ and its actual instantaneous value $\vec{M}(t)$:
$\vec{m}(t)=\vec{M}(t)-\vec{M}_0$.}
 over the film thickness $t$
and their
nonreciprocal behavior (i.e., reversal of the spinwave propagation direction causes that maximal amplitude of the dynamic magnetization reverses to the opposite interface of the FM film).
The amplitude of the dynamic magnetization of a PSSW mode over the film thickness $z$ is proportional to $\cos(m\pi z/t)$,
where the positive integer $m$ denotes the quantization number of the
standing spin wave.  

To determine the values of the exchange constant of Co$_2$MnSi,
all three dependencies of the experimental spin-wave frequencies
($\vec{H}$, $t$, $q_\|$) are compared with simulations which were
performed using a theoretical model described in detail in
Ref.~\cite{hil90}. The exchange constant $A$, the Land\'e
$g$-factor, the saturation magnetization $M_S$ and the magnetic anisotropies are the free parameters in these simulations. Because our previous BLS investigations of Co$_2$MnSi films have shown very small anisotropy in L2$_1$ ordered Co$_2$MnSi films \cite{gai08}, the anisotropy values were set to zero.

As follow from the analytical expressions~\cite{dem01,gai08,kal86}, for small spinwave wavevector used in our investigations, the frequency of the DE mode depends only marginally on the value of $A$, i.e., DE  mode frequency is particularly determined by the values of $M_S$ and $g$. On the other hand, the frequencies of the PSSW modes are particularly determined by the $A/M_S$ ratio and $g$. Furthermore, the Land\'e $g$-factor is easy to determine independently, as it scales with the slope of the BLS frequency on applied magnetic field, d$f$/d$H$. Therefore, the fitting procedure is not underdetermined and the parameters $M_S$, $A/M_S$ and $g$ are found rather independently each other, providing a high reliability (low correlation) of the fitted values.

\section{Results and discussion}

\begin{figure}
\includegraphics[width=1.0\textwidth]{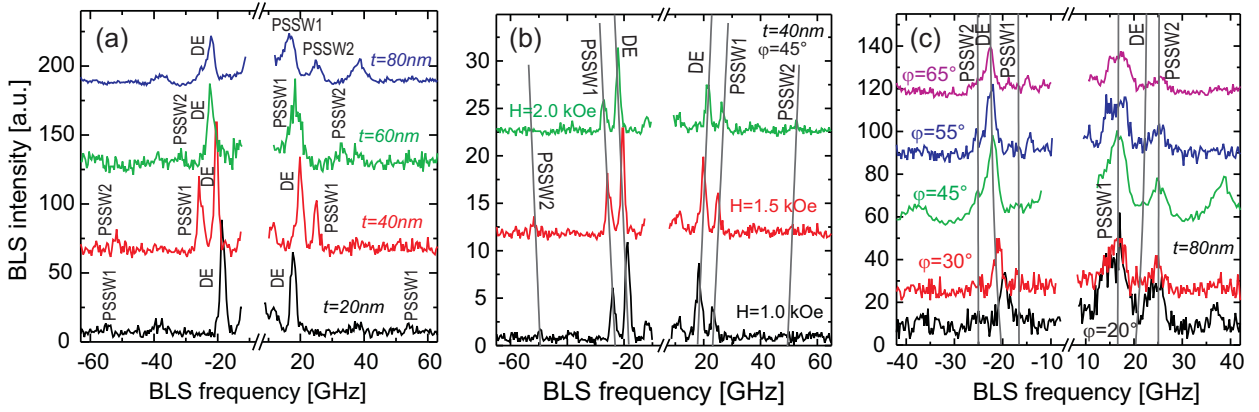} 
\caption{%
\label{figure1}%
(Color online) BLS spectra of (a)
Co$_2$MnSi films with different thicknesses $t$ acquired at an
applied external field $H=1.5$~kOe and a transferred wave vector
$q_\|=1.67\cdot10^5$~cm$^{-1}$, (b) 40\,nm thick Co$_2$MnSi film
recorded at different values of $H$ at $q_\|=1.67\cdot10^5$~cm$^{-1}$, and (c) 80\,nm thick Co$_2$MnSi film measured at $H=1.5$\,kOe and different angles of
incidence $\varphi$, i.e., different transferred wave vectors
$q_\|$. The solid line is a guide to the eye, 
showing expected peak positions as follow from the model.}
\end{figure}

Examples of BLS spectra collected from 20, 40, 60 and
\unit[80]{nm} thick Co$_2$MnSi films are shown in
Fig.~\ref{figure1}(a). The spectra were recorded at an external
magnetic field of $H=1.5$~kOe and a transferred wave vector of
$q_\|=1.67\cdot10^5$~cm$^{-1}$ (i.e.\ $\varphi=45^\circ$). 
The magnetic origin of the peaks
presented in Fig.~\ref{figure1}(a) is confirmed by 
$H$-dependent measurements,
demonstrated in Fig.~\ref{figure1}(b) for
the case of the \unit[40]{nm} thick Co$_2$MnSi film. The peak
positions in both the Stokes (negative frequencies) and the anti-Stokes
(positive frequencies) part of the spectrum move towards higher
frequencies upon increasing the field, revealing their magnetic origin. 
Figure~\ref{figure1}(c)
shows the BLS spectra recorded for the \unit[80]{nm} thick
Co$_2$MnSi film for different spin-wave wavevectors $q_\|$.
Compared to the PSSW modes, the DE mode exhibits a much stronger
dependence on the wavevector. 
Therefore, the peaks originating from the DE mode excitation can
be easily identified, whereas
 the spectral positions of the PSSW modes remain nearly
unchanged (Fig.~\ref{figure1}(c)). As expected,
the frequency of the DE mode increases with increasing film
thickness whereas the frequencies of the PSSW modes shift to lower
values (Fig.~\ref{figure1}(a)).

\begin{figure}
\includegraphics[width=0.95\textwidth]{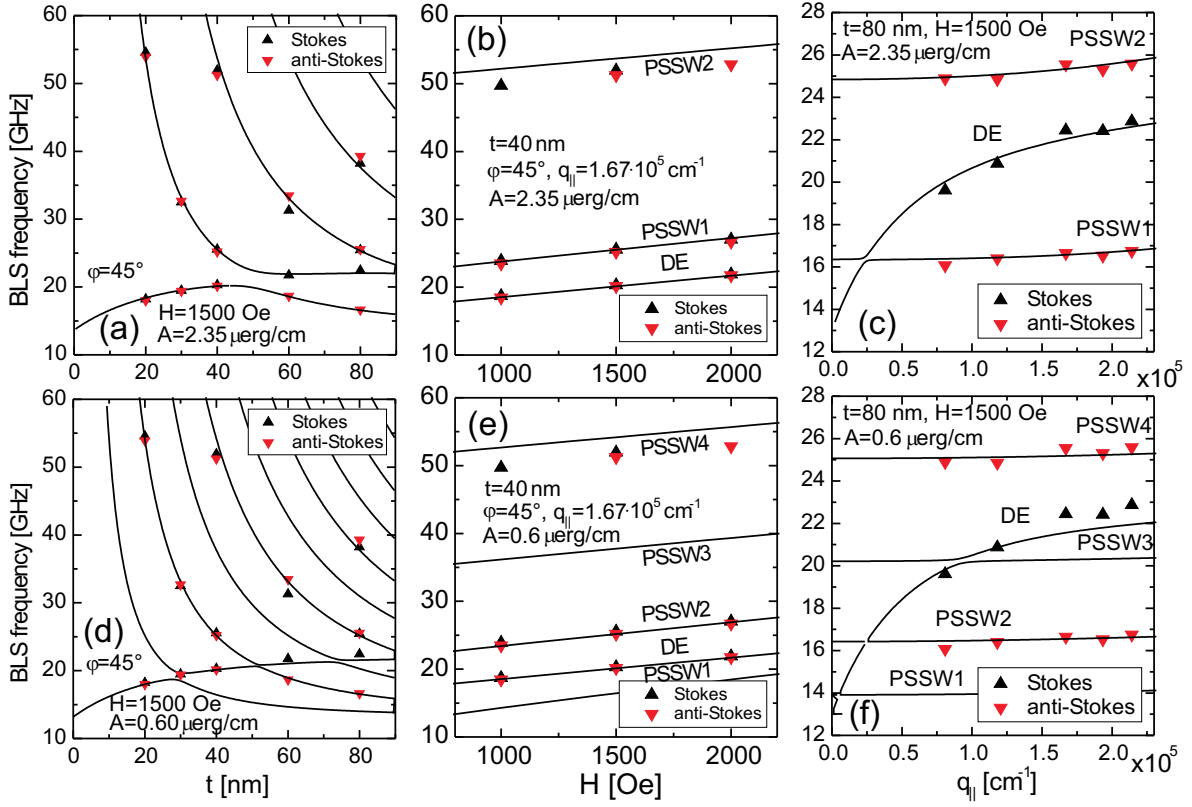}
\caption{%
\label{figA}%
(Color online) Comparison of calculated
(solid lines) and experimental (symbols) BLS frequencies as a function of (a,d) the Co$_2$MnSi thickness~$t$, (b,e) the external magnetic field~$H$ and (c,f) the transferred spin-wave wavevector $q_\|$.
Experimental data points
correspond to the BLS spectra presented in Fig.~\protect\ref{figure1}.
Triangles up (down) represent Stokes (anti-Stokes)
frequencies.
Calculation were performed using the exchange constant
(a-c) $A=2.35\pm0.1$\,$\mu$erg/cm and (d-f) $A=0.6\pm0.1$\,$\mu$erg/cm.
The remaining parameters used in the calculations are
the saturation magnetization
$M_S=970$~emu/cm$^3$ and the Land\'e $g$-factor $g=2.05$. The magnetic anisotropies are neglected.}
\end{figure}

Results of numerical simulations (solid lines) are shown in
Fig.~\ref{figA} along with the
experimentally determined BLS frequencies ($\blacktriangle$,
$\blacktriangledown$ represent Stokes, anti-Stokes frequencies,
respectively) for two different values of the exchange constant $A$. 
The saturation magnetization $M_S=970$\,emu/cm$^3$ (corresponding to $\mu=4.72$\,$\mu_B$/f.u.) and Land\'e $g$-factor $g=2.05$ is found for both values of $A$ and is in agreement with
previous investigations (e.g.\ \cite{kij06,wan05,ram06}).
An equally good agreement between the simulations and the
experimental data points is achieved for $A=2.35\pm0.1$\,$\mu$erg/cm
($D=575\pm20$\,meV\,\AA$^2$) (Fig.~\ref{figA}(a-c)) and
$A=0.60\pm0.05$\,$\mu$erg/cm ($D=145\pm10$\,meV\,\AA$^2$)
(Fig.~\ref{figA}(d-f)), respectively. When
$A=2.35\pm0.1$\,$\mu$erg/cm, the calculations describe all observed
PSSW modes. This $A$ value, however, is surprisingly large, being
nearly as large as the exchange constant reported for Co(fcc)
($A=2.73$\,$\mu$erg/cm, $D=466$\,meV\,\AA$^2$) or Co(hcp)
($A=2.85$\,$\mu$erg/cm, $D=435$\,meV\,\AA$^2$) \cite{liu96}. In
the second case where $A=0.6$\,$\mu$erg/cm, only even PSSW modes seem to be observed in the
experiment.

Obviously from the fit alone, the correct value of the exchange constant can not be obtained due to the ambiguity in mode order assignment. 
To find out which value of the exchange constant is the correct
one, we have performed analytical calculations of the BLS intensities.
In particular, we have examined under which conditions 
the BLS intensity becomes zero for odd PSSW modes only. The used
analytical approach (presented in Appendix~\ref{app:A}) 
combines an expression for the BLS intensity, as
proposed by Buchmeier \textit{et. al.}\/ \cite{buch07,buch03}, and
the analytical expression of the depth selectivity of the MOKE
effect by Hamrle \textit{et. al.} \cite{ham02}.  
The calculations
indicate that when conditions 
(\ref{eq:nullodda}) and
(\ref{eq:nulloddb}) derived in the Appendix~\ref{app:A}
are fulfilled, the BLS intensity will be zero for all odd modes. 
Using the complex refractive index of
Co$_2$MnSi $N^{(CMS)}=1.1+1.1i$ \cite{pic06}, an incidence angle
$\varphi$ of $45^\circ$ and a thickness $t$ of $20$\,nm, the value of
conditions (\ref{eq:nullodda}) and (\ref{eq:nulloddb}) becomes
$4\Re(N_z)t/\lambda=0.15$ and $4\Im(N_z)t/\lambda=0.18$, respectively. 
Those values are far from required values of an odd integer (condition
(\ref{eq:nullodda})) and zero (condition (\ref{eq:nulloddb})). This proves that
all BLS modes in our Co$_2$MnSi films 
should contribute with significant scattering cross section.
Hence the exchange
constant of Co$_2$MnSi is $A=2.35\pm0.1$\,$\mu$erg/cm (i.e.\
exchange stiffness $D=575\pm20$\,meV\,\AA$^2$), which was determined from the calculations presented in Fig.~\ref{figA}(a-c), that predicted all experimentally observed modes.

Here, we would like to comment on the values of the exchange constant
reported for Co$_2$MnSi by different groups. In Ref.~\cite{rit03},
the value of exchange stiffness was determined to be
$D=466$\,meV\,\AA$^2$ (i.e. $A=1.93$\,$\mu$erg/cm), using the
temperature dependence of the saturation magnetization. This value
is rather close to the one we have determined in our BLS investigations. In
Ref.~\cite{ram06}, the value of exchange stiffness was estimated
to be $D_B=1.9\cdot10^{-9}$Oe\,cm$^2$ (i.e. $D=225$\,meV\,\AA$^2$,
$A=0.97$\,$\mu$erg/cm) from FMR investigations, which is much
lower compared with our results. This discrepancy might originate
from the fact that in Ref.~\cite{ram06} asymmetrical pinning
conditions were assumed (i.e., the dynamic magnetization is pinned
at one interface, whereas it is unpinned at the second one). In
our samples, this condition does not apply due to the
following considerations. As can be seen in Fig.~\ref{figA}, the frequencies of the DE mode exhibit equal
values in both the Stokes and the anti-Stokes part of the spectrum. Since
the Stokes and anti-Stokes DE modes are bound to opposite 
interfaces of the FM layer, it shows that in our Co$_2$MnSi films both
interfaces are magnetically equivalent.

Finally, we would like to note that in our previous article
reporting on BLS studies of Co$_2$MnSi films \cite{gai08} the
assumption that only even PSSW modes are observed has been used.
In view of the results presented here, the conclusions of this
article remain valid. The values of the exchange constant in
Fig.~9(b) in Ref.~\cite{gai08}, however, must be scaled by a factor~5 to get the
correct values.

\section{Conclusion}

Using the case of Co$_2$MnSi thin films, we demonstrated in detail
how the values of exchange constants are determined from the BLS
measurements and carefully discussed different values of $A$
proposed by numerical simulations. The value of Co$_2$MnSi is
found to be $A=2.35\pm0.1$\,$\mu$erg/cm
($D=575\pm20$\,meV\,\AA$^2$), is agreement with the value determined 
from the temperature dependence of the magnetization \cite{rit03}.
The found value of exchange is comparable to the value of exchange constant of Co.
The pinning conditions for the dynamic
magnetization are found to be equal for both Cr/Co$_2$MnSi and
Al/Co$_2$MnSi interfaces.

\section{Acknowledgment}

The project was financially supported by the Research Unit 559
\emph{``New materials with high spin polarization''} funded by the
Deutsche Forschungsgemeinschaft, and by the Stiftung
Rheinland-Pfalz f\"ur Innovation. We thank M.~Buchmeier for
stimulating discussions.

\clearpage

\appendix
\section{Depth sensitivity of Brillouin light scattering}
\label{app:A}

The BLS intensity from a single FM layer of thickness $t$ is given by \cite{buch07}
\begin{equation}
\label{eq:IBLS0}
I^{\mathrm{(BLS)}}=I_0 \left|
\int_0^t \left[
-L(z) m_L(z) + P(z) m_p(z)
\right] \mathrm{d}z
\right|^2,
\end{equation}
where $m_L(z)$ and $m_P(z)$ are the depth profiles of the dynamic
magnetization through the FM layer in longitudinal (i.e. in the plane of the sample and
in the plane of light incidence) and polar (i.e.
out-of-plane) directions, respectively. $L(z)$, $P(z)$ are complex
depth sensitivity functions of the off-diagonal reflectivity
coefficient $r_{sp}$ for longitudinal $M_L(z)$ and polar $M_P(z)$ static
magnetization profiles \cite{buch07, ham02}
\begin{equation}
r_{sp}(\vec{M})=
r_{sp}(M\!\!=\!\!0)+\int_0^t \left[
L(z) M_L(z) + P(z) M_P(z)
\right] \mathrm{d}z.
\end{equation}
The dependence of $L(z)$, $P(z)$ on depth $z$ is the same for both
of them. Assuming that the substrate has a refractive index identical to the refractive
index of the FM layer,  the analytical terms of $L(z)$ and $P(z)$  may be written as
\cite{ham02}
\begin{align}
\label{eq:Lz}
L(z)&=L(0)\exp[-4\pi i N_z z/\lambda] \\
\label{eq:Pz}
P(z)&=\gamma L(z),
\end{align}
where the complex coefficient $\gamma$ is the ratio of $P(0)$
and $L(0)$. $N_z$ is the normalized wave vector in polar direction
$N_z=\sqrt{(N^\mathrm{(fm)})^2 -
(N^\mathrm{(air)})^2\sin^2\varphi}$, where $N^\mathrm{(fm)}$ and
$N^\mathrm{(air)}$ are the refractive indices of the FM layer and air,
respectively, and $\varphi$ is the angle of incidence of the
probing light beam with respect to the sample's normal. 
Note that if the refractive index of the substrate
is different from the one of the FM layer, the analytical expressions
become more complex. However the basic features (i.e.\ continuous
decay of the amplitude and continuous shift of the phase)
remain valid.

Using the open (anti-pinning) boundary conditions at the interfaces of the FM layer, the depth profile of the dynamic magnetization of a PSSW mode of the
$m$-th order reads \cite{kal86}
\begin{align}
  \label{eq:ml}
  m_L(z,\tau)&=m_0 \cos(m\pi z/t) \cos(\omega_{sw}\tau) \\
  \label{eq:mp}
  m_P(z,\tau)&=m_0 \varepsilon \cos(m\pi z/t) \cos(\omega_{sw}\tau+\pi/2),
\end{align}
where $\omega_{sw}$ and $\tau$ are the frequency of the spin-wave mode and time, respectively. The magnetization vector follows an elliptical trajectory, with ellipticity $\varepsilon$. Therefore, the $m_L$ and $m_P$
magnetization components are shifted by $\pi/2$ in their time
dependence.

Combining Eqs.(\ref{eq:IBLS0}, \ref{eq:Lz} -- \ref{eq:mp}) and
integrating over the thickness of the FM layer and averaging over
time, we obtain
\begin{equation}
\label{eq:IBLS}
I^\mathrm{(BLS)}=
\frac{1}{2} I_0
(1+| \varepsilon \gamma |^2 )
\left|
\frac{L(0) \alpha t m_0}%
 { m^2\pi^2 - \alpha^2}
\right |^2
\left|
1-\exp(-i\alpha) (-1)^m
\right|^2,
\end{equation}
with $\alpha$ being a dimensionless parameter defined as $\alpha=4\pi N_z t/\lambda$.

The total BLS intensity $I^\mathrm{(BLS)}$ is zero when the last
term in Eq.(\ref{eq:IBLS}) is zero. For odd $m$, the intensity is zero when the following conditions hold 
\begin{align}
\label{eq:nullodda}
4 \Re(N_z) t/\lambda&=(2k+1) \\
\label{eq:nulloddb}
4 \Im(N_z) t/\lambda &= 0,
\end{align}
where $k$ is integer. Note that for $m=2k+1$,
the denominator in Eq.~(\ref{eq:IBLS}) becomes zero as well, 
which would also provide a zero BLS intensity. For even $m$,
$I^\mathrm{(BLS)}$ is zero when
\begin{align}
2  \Re(N_z) t/\lambda&=k  \\
4 \Im(N_z) t/\lambda &= 0.
\end{align}
where $k$ is again integer.

\section{Units of exchange}
\label{app:B}

Throughout the literature, the exchange is expressed in at least four
different variables, making the direct comparison of exchange
values difficult. This situation is further complicated by the use of
both cgs and SI units, as well as the lack of a unique
nomenclature. Here, we give a short overview of different
definitions of exchange and the conversion between them.

The energy of a single magnon (i.e. a spin wave, where one single
spin in one period of spin wave is reversed) having wave vector
$k$ is given by \cite{siegmann}
\begin{equation}
\label{eq:D} E=\hbar\omega=D k^2,
\end{equation}
where $D$ is called the exchange stiffness (which is sometimes 
refereed to as the spin-wave
stiffness), $\omega$ is the angular frequency of the spin wave, and
$\hbar$ is the reduced Planck constant. The routinely used units of $D$ are
meV\,\AA$^2$ for both cgs and SI units.

Alternatively, the exchange is expressed as the strength of the
effective field $D_B$ created by the exchange energy of the
spin wave. As $D$, $D_B$ is also often called the exchange
stiffness. Neglecting a dipolar contribution to the spin wave (i.e.\
assuming a large value of $k$), the angular frequency of a spin wave is given
by
\begin{align}
\label{eq:DB} \frac{\omega}{g \gamma_0} & =(B+D_B k^2)
\end{align}
where $\omega$ is the angular frequency of the spin wave, $g$ is
Land\'e $g$-factor, $B$ is an external field and $g\gamma_0$ is the gyromagnetic ratio, where $\gamma_0=e/(2m_e)=8.793\cdot10^{10}$\,s$^{-1}$T$^{-1}$ for SI and $\gamma_0=e/(2cm_e)=8.793\cdot10^{6}$\,s$^{-1}$G$^{-1}$ for cgs, $c$, $e$ and $m_e$ being the speed of the light, the charge and the rest mass of an
electron, respectively. Units of $D_B$ are Oe cm$^2$ in cgs, and
T\,m$^2$ or J\,A$^{-1}$ in SI.

Another description of the exchange is provided by the exchange constant
$A$ (sometimes also called the exchange stiffness). Within this
description, the energy density $E/V=\mathcal{E}$ of a given
continuous magnetization distribution $\vec{m}(\vec{r})$ within
space is given by
\begin{equation}
\label{eq:Adef} E/V= \mathcal{E} =A |\nabla \vec{m}| ^2,
\end{equation}
where $\vec{m}=\vec{M}/M_S$ is the reduced magnetization and $M_S$
is the saturation magnetization. Units of $A$ are erg/cm in cgs and
J/m in SI. Assuming a reduced magnetization $\vec{m}$ in form of a spin
wave with wavevector $k$, the effective field of such a spin wave
is
\begin{equation}
\label{eq:Beff} B_\mathrm{eff}=-\nabla_M \mathcal{E}= \frac{2A}{M_S}
k^2.
\end{equation}

Comparing Eqs. (\ref{eq:D}--\ref{eq:Beff}), the conversion between
$D$, $D_B$ and $A$ are (valid both in SI and cgs)
\begin{align}
D_B& =2A/M_S, \\
D_B& =\frac{D}{g \gamma_0 \hbar}\equiv\frac{D}{g\mu_B} ,\\
A& =\frac{D M_S}{2 g \gamma_0 \hbar}\equiv\frac{D M_S}{2g\mu_B},
\end{align}
where $\mu_B=\gamma_0\hbar$ is the Bohr magneton.
Note that the conversion between SI and cgs for the exchange constant
is given by: 1$\,$erg/cm=$10^{-5}$\,J/m. For the saturation magnetization
1\,emu/cm$^3$=1000\,A/m. If $M_S=1$\,emu/cm$^3$, then $4\pi
M_S=4\pi$ G (i.e.\ in unit of Gauss).

Finally, here we mention a description of exchange coming from the effective Heisenberg
Hamiltonian with classical spins
\begin{equation}
\label{eq:Jij}
E=-\sum_{i\neq j} J_{ij} \vec{s}_i \cdot \vec{s}_j,
\end{equation} 
where $J_{ij}$ is the exchange
interaction energy between two spins at positions $i$, $j$ and $s_i$ , $s_j$ are 
unit vectors pointing in the direction of local magnetic moments at sites $i$, $j$, respectively.
Assuming that a unit cell contain a single atom, then the spin-wave energy $E(\vec{k})$ is related to the exchange 
parameters $J_{ij}$ by a simple Fourier transformation \cite{paj01}
\begin{equation}
\label{eq:Jij2D}
E(\vec{k})=\frac{4\mu_B}{\mu} \sum_{j\neq 0} J_{0j} 
\left(1-\exp(i\vec{k}\cdot\vec{R}_{0j})
\right),
\end{equation}
where $\vec{R}_{0j} =\vec{R}_0 -\vec{R}_j$ denotes a lattice vector in real space, $\vec{k}$ is the spinwave vector, $\mu$ is the magnetic moment per atom and $\mu_B$ is the Bohr magneton. The exchange stiffness $D$ is given by the curvature of the spin-wave dispersion $E(\vec{k})$ at  $\vec{k} = 0$. When unit cell contains several atoms, the spin-wave energy is expressed e.g.\ in Ref.~\cite{kub07}.


\begin{thebibliography}{10}

\bibitem{hillebrands07}
Hillebrands B, Hamrle J.
\newblock Investigation of Spin Waves and Spin Dynamics by Optical Techniques.
\newblock In: Handbook of Magnetism and Advanced Magnetic Materials Weinheim:
  Wiley-Interscience; 2007.

\bibitem{liu96}
Liu X, Steiner MM, Sooryakumar R, Prinz GA, Farrow RFC, Harp G 1996 {\em Phys.
  Rev. B} {\bf 53} 12166

\bibitem{buch03}
Buchmeier M, Kuanr BK, Gareev RR, B\"urgler DE, Gr\"unberg P 2003 {\em Phys.
  Rev. B} {\bf 67}(18) 184404

\bibitem{moc87}
Mock R, Hillebrands B, Sandercock R 1987 {\em J. Phys. E: Sci. Instrum.} {\bf
  20} 656

\bibitem{hil99}
Hillebrands B 1999 {\em Rev. Scien. Instr.} {\bf 70} 1589

\bibitem{gai08}
Gaier O, Hamrle J, Hermsdoerfer SJ, Schulthei\ss\ H, Hillebrands B, Sakuraba Y,
  Oogane M, Ando Y 2008 {\em J. Appl. Phys.} {\bf 103}(10) 103910

\bibitem{dem01}
Demokritov SO, Hillebrands B, Slavin AN 2001 {\em Physics Reports} {\bf 348}
  441

\bibitem{kal86}
Kalinikos BA, Slavin AN 1986 {\em J. Phys. C} {\bf 19} 7013

\bibitem{hil90}
Hillebrands B 1990 {\em Phys. Rev. B} {\bf 41} 530

\bibitem{ram06}
Rameev B, Yildiz F, Kazan S, Aktas B, Gupta A, Tagirov LR, Rata D, Buergler D,
  Gr\"unberg P, Schneider CM, K\"ammerer S, Reiss G, H\"utten A 2006 {\em Phys.
  Stat. Sol. (a)} {\bf 203} 1503

\bibitem{kij06}
Kijima H, Ishikawa T, Marukame T, Koyama H, Matsuda K, Uemura T, Yamamoto M
  2006 {\em IEEE Trans. Mag.} {\bf 42} 2688

\bibitem{wan05}
Wang WH, Przybylski M, Kuch W, Chelaru LI, Wang J, Lu YF, Barthel J, Meyerheim
  HL, Kirschner J 2005 {\em Phys. Rev. B} {\bf 71}(14) 144416

\bibitem{buch07}
Buchmeier M, Dassow H, B\"{u}rgler DE, Schneider CM 2007 {\em Phys. Rev. B}
  {\bf 75}(18) 184436

\bibitem{ham02}
Hamrle J, Ferr\'e J, N\'yvlt M, Vi\v{s}\v{n}ovsk\'y v 2002 {\em Phys. Rev. B}
  {\bf 66}(22) 224423

\bibitem{pic06}
Picozzi S, Continenza A, Freeman AJ 2006 {\em J. Phys. D: Appl. Phys.} {\bf
  39}(5) 851

\bibitem{rit03}
Ritchie L, Xiao G, Ji Y, Chen TY, Chien CL, Zhang M, Chen J, Liu Z, Wu G, Zhang
  XX 2003 {\em Phys. Rev. B} {\bf 68}(10) 104430

\bibitem{siegmann}
St\"ohr J, Siegmann HC.
\newblock Magnetism. From fundamentals to nanoscale dynamics.
\newblock Springer; 2006

\bibitem{paj01}
Pajda M, Kudrnovsk\'y J, Turek I, Drchal V, Bruno P 2001 {\em Phys. Rev. B}
  {\bf 64}(17) 174402

\bibitem{kub07}
K\"{u}bler J, Fecher GH, Felser C 2007 {\em Phys. Rev. B} {\bf 76}(2) 024414

\end{thebibliography}

\end{document}